# Spin-Orbit-Torque-based Devices, Circuits and Architectures

Farshad Moradi, *Senior Member*, Hooman Farkhani, *Member, IEEE, IEEE*, Behzad Zeinali, *Member, IEEE*, Hamdam Ghanatian, Johan Michel Alain Pelloux-Prayer, *Member, IEEE*, Tim Boehnert, *Member, IEEE*, Mohammad Zahedinejad, Hadi Heidari, Senior Member, IEEE, Vahid Nabaei, Member, IEEE, Ricardo Ferreira, Johan Akerman, Jens Kargaard Madsen, *Member, IEEE*

*Abstract*—**Spintronics, the use of spin of an electron instead of its charge, has received huge attention from research communities for different applications including memory, interconnects, logic implementation, neuromorphic computing, and many other applications. Here, in this paper, we review the works within spintronics, more specifically on spin-orbit torque (SOT) within different research groups. We also provide researchers an insight into the future potentials of the SOT-based designs. This comprehensive review paper covers different aspects of SOT-based design from device and circuit to architecture level as well as more ambitious and futuristic applications of such technology.**

## I. INTRODUCTION

TODAY, spintronics has found its footprint on many applications ranging from more commercialized ones such as spin-based non-volatile memories, e.g. Spin-Transfer Torque Magnetic Random Access Memory (STT-MRAM) or SOT-MRAMs [1]-[4] and magnetic sensors for sensing magnetic fields in the range of picoTesla [5], to more emerging technologies such as neuromorphic [6] and non-Boolean computing [7], [8], interconnects [9] and logic implementations [10]. Spintronics is based on contribution of both electron charges and electron spin to enable the electronic functionalities, giving one extra degree of freedom for computing purposes. Historically, the field was initially pushed by sensing applications with a focus on the magnetic storage industry and automation control, and more recently, a big effort has been put in the development of ultra-dense non-volatile memories, which is mainly due to the fact that spintronic devices offer unique features including non-volatility, high endurance and CMOS-compatibility. Such features are necessary for future memories mainly due to the fact that conventional memories such as Static Random-Access Memories (SRAMs), Dynamic Random-Access Memories (DRAMs), etc. have faced low yield and high leakage currents. These issues are mainly attributed to the miniaturization of the Complementary Metal–Oxide–Semiconductor (CMOS) technology beyond 22nm leading to significantly increase in process variations, short channel effects (SCE) and leakage [11]. Although different solutions at device, circuit, and architecture levels have been developed to relax leakage power and unreliability of volatile memories [12]-[14], non-volatile memories using emerging technologies have shown a high potential to alleviate or even eliminate such issues. Among different types of non-volatile memories, MRAMs developed based on spintronics have attracted huge attention.

The storage element in spintronic MRAMs is a Magnetic Tunnel Junction (MTJ) shown in Fig. 1(a). The MTJ is composed of two magnetic layers including a pinned layer (PL) and a free layer (FL), and one oxide barrier layer (e.g. MgO). This configuration gives two permanent states with different resistance values in the MTJ. MTJ resistance is determined by the relative magnetization directions of FL and PL. When the magnetization directions of the two magnetic layers are parallel (P), the MTJ resistance is low. For the anti-parallel (AP) state, the resistance is high. Therefore, the MTJ is able to store binary bits, low resistance ($R_L$) as logic "0" and high resistance ($R_H$) as logic "1".

The analog character of the MTJ can be used for different applications. Such example is the use of MTJ as a tunable oscillator (Fig. 1 (b) and (c)) [15], [16]. Such spintronic-based oscillator is extremely compact (more than 50X smaller than a standard Voltage Controlled Oscillator (VCO) designed in a CMOS process) with a wide tuning range, which can reach 100% for carrier frequencies that are in the range of 1 to 10 GHz and higher [15], [16]. In practice, to distinguish between switching and oscillation, a critical current density ($J_c$) can be defined; when the generated spin current inside an MTJ resulting from external forces is bigger than $J_c$, the switching can be enabled. On the other hand, when the spin current density is smaller than $J_c$, the MTJ may oscillate. The technology trend is to reduce $J_c$ making MTJs appropriate for low-power applications. However, it is challenging to scale $J_c$, while keeping thermal stability ($\Delta$) high to fulfil the stability of the circuit at the presence of thermal fluctuations. In other words, easier demagnetization (lower $J_c$) can be achieved at the cost of lower stability (smaller $\Delta$). To alleviate this tradeoff, different methods have been developed [17]-[20]. The most practical solution is to change the easy axis direction of the

Farshad Moradi, Hooman Farkhani, Johan Michel Alain Pelloux-Prayer and Jens Kargaard Madsen are with Integrated Circuits and Electronics Lab (ICELab) at Aarhus University, Denmark.
Behzad Zeinali and Hamdam Ghanatian are independent researchers.
Vahid Nabaei and Hadi Heidari are with MeLab at University of Glasgow, UK.

Tim Boehnert and Ricardo Ferreira are with the INL, Portugal.
Mohammad Zahedinejad and Johan Akerman are with the University of Gothenburg, Sweden.
Corresponding author: moradi@eng.au.dk

MTJ. In in-plane-anisotropy MTJ (i-MTJ), the easy axis is in $x$-axis direction. The thermal stability of i-MTJs is limited by the shape of the MTJ due to the requirement of an elliptical shape to stabilize the magnetization along the in-plane axis in order to minimize the magnetostatics energy. However, keeping the elliptical shape and preventing magnetization curling are almost impossible in scaled technology nodes [20]. When an in-plane magnetic film is patterned on a submicron scale, magnetization curling occurs at the edges of the film, leading to increased dispersion and unreliability of the switching characteristics of the MTJ [21]. To alleviate such effect as a rule of thumb, the aspect ratio of the MTJ, i.e. length to width, must be bigger than two that sacrifices the density of the memory [21]. To deal with this, perpendicular-anisotropy MTJs (p-MTJ) in which the easy axis is aligned in the $z$-axis direction can be used. The p-MTJ alleviates the size and shape limitations leading to a higher memory density with higher thermal stability, and they may present lower $J_c$ and higher switching efficiency [20].

Several methods have been developed to control the magnetization of the FL (either switching or oscillation) [22]-[24]. The main mechanism requires only a bidirectional current to apply a STT on the magnetization of the MTJ as shown in Fig. 1(d). The bidirectional current ($I_{STT}$) is generated by an appropriate bias of the bit-line (BL) and source-line (SL), and asserting the word-line (WL) signal meaning that electrons may enter the MTJ through either PL or FL. When unpolarized electrons enter from the PL side, their spin aligns with that of the PL's, and after tunneling into the FL, their spin momentum is conserved and injected into the FL. If the injected moment is large enough, the magnetization direction of the FL aligns with PL's magnetization and MTJ is switched to P-state. On the other hand, when unpolarized electrons enter from the FL side, electrons with the same spin direction as PL will be able to tunnel across the oxide barrier easily while electrons with the opposite spin direction accumulate in the FL. This spin accumulation in the FL exerts a torque on FL that favors the orientation opposite to the PL magnetization. Thus, if the starting configuration is parallel and the spin accumulation is large enough, it can force the magnetization direction of the FL to switch to the opposite direction of the PL, and the MTJ is switched to the AP-state. The generated STT is proportional to $\overrightarrow{m_{FL}} \times \overrightarrow{m_{FL}} \times \overrightarrow{m_{PL}}$. In order to model the time evolution of the magnetization direction of the FL during the switching process, Landau-Lifshitz-Gilbert (LLG) equation is usually employed [20], [25], [26]. When the FL is demagnetized by using the STT mechanism, LLG equation can be expressed as follows:

$$\frac{\partial \overrightarrow{m_{FL}}}{\partial t} = -\gamma \mu_0 \overrightarrow{m_{FL}} \times \vec{H}_{eff} + \alpha \overrightarrow{m_{FL}} \times \frac{\partial \overrightarrow{m_{FL}}}{\partial t} - \frac{\gamma h}{2qt_{FL}M_s} \frac{pI_{STT}}{A_{MTJ}} \overrightarrow{m_{FL}} \times \overrightarrow{m_{PL}} \quad (1)$$

where, $\overrightarrow{m_{FL}}$ describes the magnetization vector of the FL containing three elements of $m_x$, $m_y$ and $m_z$ along the $x$-, $y$- and $z$-axis, respectively. The first two terms in the right side of the LLG equation represent the field-induced and Gilbert damping torques, which describe the precession of the magnetic moment around the effective magnetic field ($H_{eff}$ contains anisotropy and demagnetization fields) and the relaxation of the moment until it is aligned with the effective field, respectively. In these terms, $\gamma$ and $\alpha$ are the gyromagnetic ratio and Gilbert damping constant, respectively. In the absence of any external demagnetization force (in this case $I_{STT}$), the magnetization direction of the FL is forced to be aligned with the easy axis. When the magnetization direction of the FL is not collinear with the easy axis, anisotropy field pulls the magnetization direction of the FL back to the easy axis. STT is modeled by the last item in the right side of the equation, where $M_s$, $p$ and $\overrightarrow{m_{PL}}$ represent the saturation magnetization, effective spin polarization and the PL magnetization vector, respectively. In addition, the physical parameters $t_{FL}$ and $A_{MTJ}$ are the FL thickness and the cross-sectional area of the MTJ, respectively. Depending on the sign of $I_{STT}$, STT either increases or decreases the magnitude of the precession (the first term in the right side of (1)). To capture the MTJ resistance ($R_{MTJ}$) during the evolution of $\overrightarrow{m_{FL}}$, the non-equilibrium Green's function (NEGF) approach is utilized leading to the following equation [27]:

$$R_{MTJ} = R_{P0} \frac{1 + \left(\frac{V_{MTJ}}{V_h}\right)^2 + TMR_0}{1 + \left(\frac{V_{MTJ}}{V_h}\right)^2 + TMR_0\left[\frac{1 + \cos\theta}{2}\right]} \quad (2)$$

where $\theta$ is the polar angle (angle between $\overrightarrow{m_{FL}}$ and the $z$-axis). It is worth mentioning that $\overrightarrow{m_{FL}}$ can be completely characterized by $\theta$ and $\varphi$; where $\varphi$ is the azimuthal angle (angle between the orthogonal projection of $\overrightarrow{m_{FL}}$ on the $x$-$y$ plane and the $x$-axis). In NEGF, $R_{p0}$ and $TMR_0$ are $R_p$ and $TMR$ at zero bias voltage while $V_h$ is the voltage at which $TMR_0$ is divided by 2.

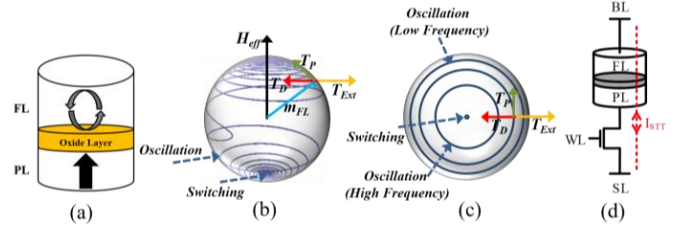

Fig. 1. (a) MTJ, the evaluation of $m_{FL}$ in switching and oscillation processes. (b) side view, (c) top view. $H_{eff}$ is the effective magnetic field applied to $m_{FL}$. $T_P$ describes the precession torque leading to demagnetization of $m_{FL}$. $T_D$ is the damping torque that aligns $m_{FL}$ with $H_{eff}$ and $T_{Ext}$ is the torque induced by external forces. This torque pushes $m_{FL}$ to either switch to $\pm z$ or oscillate with a specific frequency. (d) 1T1J STT-MRAM cell.

STT technology suffers from unreliability in scaled technology nodes as by scaling the size and area of the MTJ, the oxide layer must be scaled down in order to maintain a reasonable device resistance that allows $J_c$ to be reached within the dielectric breakdown limit of the tunnel barrier. This is particularly challenging for small technology nodes, which require the tunnel barriers density to approach thicknesses of the order of 0.5 nm. At this level, the insulating tunnel barrier becomes leaky with tunneling current channels co-existing with metallic transport channels that do not conserve the moment. As a result, the spin polarization decreases, and so does the current density required to switch the memory elements.

The mentioned issue in STT-based devices can be addressed by SOT (Fig. 2(a)) [28]-[30]. SOT-based dynamics are induced by a charge current ($I_{SOT}$) that must flow through a non-magnetic heavy metal (HM) adjacent to the FL. This is a very important difference, since now the electrical current required to excite dynamics is no longer pushed through the thin tunnel barrier. For STT, the free layer dynamics have their origin in the injection of a spin polarized current in the free layer while for SOT, this spin polarized current is created by either the Rashba effect [31] or the Spin-Hall effect [32], [33]. The Rashba effect originates from breaking of structural symmetry [31] as shown in Fig. 2(b). Structural asymmetry in magnetic heterostructures generates an interaction between electrical current and magnetic resulting in a torque exerted on the magnetic material. In Fig. 2(a), the FL is sandwiched between two different materials (HM and oxide layer), hence the vertical structural symmetry is broken. Therefore, passing $I_{SOT}$ generates a torque interacting on the magnetization direction of the adjacent FL. Such torque can be characterized by the Rashba magnetic field ($\vec{H}_R$):

$$\vec{H}_R = \frac{\alpha_R}{A_{HM}}\left(\vec{a}_z \times \vec{I}_{SOT}\right) \qquad (3)$$

where $\alpha_R$ is the Rashba coefficient, $A_{HM}$ is cross-sectional area of the HM and $\vec{a}_z$ is the unit vector along the $z$-axis. SOT induced by Rashba magnetic field is proportional with $\overrightarrow{m_{FL}} \times \vec{H}_R$. According to (3), by flowing the charge current along the $y$-axis, an effective magnetic field in the $x$ direction will be generated. The SOT can be also justified by the SHE translated as the ability of generating a spin current from a non-spin-polarized charge current. In SHE as shown in Fig. 2(c) and (d), $I_{SOT}$ can generate spin accumulation on the lateral surfaces of the HM due to the strong spin-orbit interaction, which forms a pure spin current ($I_s$). $I_s$ is orthogonal to both $I_{SOT}$ and electron spin [30]and may be characterized as follows:

$$\vec{I}_s = \frac{A_{MTJ}}{A_{HM}}\eta\left(\vec{I}_{SOT} \times \vec{\sigma}_{SHE}\right) \qquad (4)$$

where $\vec{\sigma}_{SHE}$ and $\eta$ are the polarization direction of the spin current and the spin Hall angle, respectively. $I_s$ is injected into the adjacent FL generating a SOT proportional to $\overrightarrow{m_{FL}} \times \overrightarrow{m_{FL}} \times \vec{\sigma}_{SHE}$. According to (4), effective spin injection efficiency can be defined as $\frac{I_s}{I_{SOT}} = \frac{\eta A_{MTJ}}{A_{HM}}$ which may be larger than 100% if $A_{HM}$ is designed much smaller than $A_{MTJ}$. The origin of the interaction leading to SOT is not completely specified; it may be possibly due to the injection of a spin current from the HM into the FL of the MTJ induced by SHE and/or a field-like torque on the FL due to the Rashba effect. However, the SOT induced by the Rashba effect is more subjected to debate [27]. Therefore, for simplicity, we consider that SHE is the driver of the SOT. Therefore, LLG equation in (1) is modified by adding the torques generated by $I_{SOT}$ and an external magnetic field ($H_{ext}$). When a p-MTJ is demagnetized by SOT mechanism, adding an $H_{ext}$ is necessary for deterministic demagnetization (Section II). In the modified LLG, the torque induced by the injected spin current is modeled similarly to that of STT:

$$\frac{\partial \overrightarrow{m_{FL}}}{\partial t} = -\gamma\mu_0 \overrightarrow{m_{FL}} \times \left(\vec{H}_{eff} + \vec{H}_{ext}\right) + \alpha\overrightarrow{m_{FL}} \times \frac{\partial \overrightarrow{m_{FL}}}{\partial t} - \frac{\gamma h}{2qt_{FL}M_s}\left[\frac{pI_{STT}}{A_{MTJ}}\overrightarrow{m_{FL}} \times \overrightarrow{m_{FL}} \times \vec{m}_{PL} + \frac{\eta I_{SOT}}{A_{HM}}\overrightarrow{m_{FL}} \times \overrightarrow{m_{FL}} \times \vec{\sigma}_{SHE}\right] \qquad (5)$$

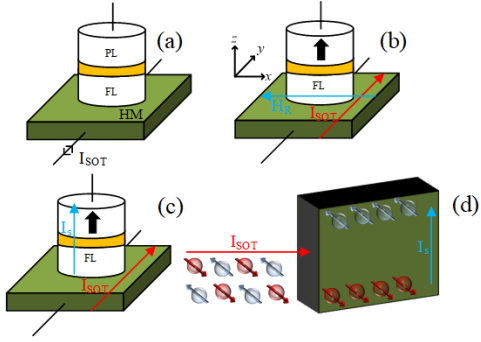

Fig. 2. (a) SOT-based MTJ. (b) Rashba Magnetic field ($H_R$) in the MTJ. (c) Spin current ($I_s$) generated by SHE in the MTJ. (d) Cross-sectional view of the HM showing the SHE; Unspin-polarized charge current ($I_{SOT}$) enters to the HM, then the HM segregates spins and generates a spin current ($I_s$).

Based on the described advantages of SOT in comparison with the STT, there is a strong demand to use this technology in emerging applications of spintronics. Most of the applications of spintronics use either *Switching* or *Oscillation* of the MTJ.

In this paper, in section II, the advantages and disadvantages of using SOT technology in memory applications will be explained. The MRAMs developed based on SOT technology (SOT-MRAM) will be reviewed with their entailed challenges, and different approaches to make them more practical will be introduced. In section III, we will give a comparison between SOT-MRAM and other spin-based memories. In section IV, the application of SOT devices as oscillator for radio frequency applications will be discussed. Afterwards, in section V, the emerging applications of SOT in artificial intelligence (AI) and computation will be reviewed. In this section, the SOT-based neuromorphic systems and approximate computing systems will be investigated. In addition, the possible improvements that may be achieved by embedding photonics and spintronics will be elaborated. Section VI will investigate another interesting application of SOT technology in non-Boolean computing systems. In this section, two critical building blocks of such computing systems, e.g. Analog-to-Digital converter and analog sensors implemented by SOT technology will be introduced. This application can be considered as a hot topic for future research activities. In section VII, other emerging applications of SOT devices such as wireless communication systems and rain interfacing systems will be discussed. The future vision and existing challenges of SOT-based applications (e.g. integration and CMOS-interfacing circuit design) will be explored in section VIII. Finally, the paper will be concluded in section IX.

## II. MEMORY DESIGN APPLICATIONS

As mentioned in Section I, spintronic devices have been developed primarily for memory applications. In such applications, a MTJ is used as a storage element to store a binary data. To write the binary data, $I_{SOT}$ must be kept long enough to make sure that the switching process is accomplished. This is mainly due to stochastic switching of the MTJ, which will be described further in the following subsections. On the other hand, to read the stored data, the resistive feature of the MTJ is utilized. To this end, a small sensing current passes through the MTJ and the voltage drop across the MTJ is measured. To enable a reliable read operation, the sensing current must be small enough to prevent data disturbance. On the other hand, having a high read margin is necessary for reliable read operation enabled by high sensing current [34]. The read margin is defined as the difference between the voltage drop across the MTJ and the reference voltage used for sensing. A high sensing current distinguishes further the two states of the voltage drop, hence, the probability of read failure decreases. Therefore, it is challenging to keep both read failure and disturbance probabilities low.

Here, different SOT-based memory structures will be reviewed. First, the conventional SOT-MRAM cell including its write/read mechanisms is explained. Then, the deterministic switching using p-MTJs will be explained. Multilevel SOT-MRAM cells for implementation of more compact memory arrays will be presented. Finally, assisting techniques by which specifications of the SOT-based memories can be improved, will be presented.

### A. SOT-MRAM

Figure 3(a) shows the the conventional SOT-MRAM consisting of one SOT-based MTJ and two access transistors ($M_{write}$ and $M_{read}$). In addition, there are two access ports called source-line (SL) and bit-line (BL) as well as two word-line (WL) signals: one for read operation (RWL) and the other for write operation (WWL). During write operation, either BL or SL is connected to $V_{DD}$ while the other one is grounded. The biasing depends on the binary data to be written into the MTJ. Afterwards, WWL is connected to $V_{DD}$ while RWL is grounded, which will cause the $M_{write}$ to turn on and $M_{read}$ to turn off. Therefore, $I_{SOT}$ will flow from either BL to SL or vice versa. Since $M_{read}$ is off, almost no current leaks into the MTJ. WWL must be kept high until the switching is finished. During read, SL and WWL are grounded while RWL is connected to $V_{DD}$. This configuration leads to passing a sense current ($I_{sense}$) through $M_{read}$ and generating a BL voltage (i.e. $V_{BL}$). Afterwards, the resistance state of the MTJ can be measured using the conventional sensing circuit shown in Fig. 3(b). In this circuit, the stored data is read out by comparing the MTJ resistance with a reference value [35]. $I_{sense}$ is generated in the reference branch and conveyed to the data branch through a current mirror circuit. The value of $I_{sense}$ can be tuned by a clamp

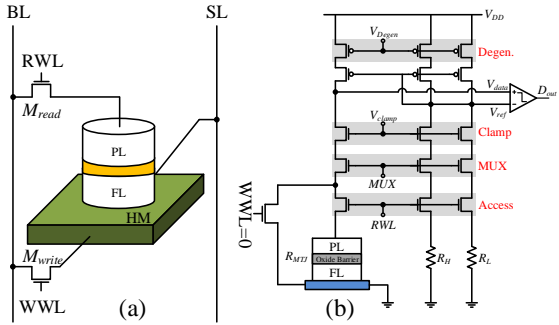

Fig. 3. Conventional (a) SOT-MRAM cell and (b) sensing circuit

voltage ($V_{clamp}$) biasing clamp NMOS transistors. In addition,

degeneration resistances are employed in the source of the clamp transistors reducing process variation in the sensing circuit. Degeneration PMOS transistors are added in the circuit to provide a constant current passing through the current mirror circuit. To improve read margin, the reference voltage ($V_{ref}$) is selected between $V_{BL,H}$ and $V_{BL,L}$ defined as the BL voltage when the MTJ is at high and low resistance states, respectively.

The above description shows that read and write operations are enabled by activation of separate access transistors. Therefore, in such memory cell, read and write operations can be independently optimized. However this advantage comes with a penalty of area overhead of an extra transistor in each cell (30% overhead in [36]) makes the usability of such SOT-MRAMs questionable in ultra-compact memory applications.

One approach to deal with this issue is presented in [36] in which $M_{read}$ is replaced by a Schottky diode. This idea is inspired from the fact that during read operation, $I_{sense}$ always flows from BL to SL. Therefore, we need to provide a current path during read operation while during write operation this path should be disconnected. This feature can be obtained by a diode; the Schottky diode is forward biased during read, whereas it is reverse biased during a write. This structure may solve the area overhead issue at the cost of more complicated fabrication process, although this depends on the implementation of the diode in a specific technology; area overhead improvement can be only few percent.

### B. Deterministic p-MTJ SOT-MRAM

As mentioned earlier, WWL must be kept active long enough to accomplish a switching. It is mainly due to the inherent stochastic switching process of the MTJ. Total switching time of the MTJ is a combination of *incubation* and *transit* times [37], [38]. The total switching time is mainly determined by the incubation time, hence stochastic behavior of the switching time results from variations in the incubation time [38].

According to the LLG equation (1), theoretically, initial STT term is zero because the magnetization of the FL and PL are exactly collinear, hence $\overrightarrow{m_{FL}} \times \overrightarrow{m_{PL}} = 0$. However, in a real situation, thermal fluctuation creates a small misalignment between $\overrightarrow{m_{FL}}$ and $\overrightarrow{m_{PL}}$ enabling a small STT to trigger the switching process. Therefore, if the STT current is large enough, $\overrightarrow{m_{FL}} \times \overrightarrow{m_{PL}}$ evolves slowly during a long time, the so-called incubation delay. In the SOT-based i-MTJ technology, the switching time is still dominated by the incubation delay mainly because $\overrightarrow{m_{FL}} \times \overrightarrow{\sigma_{SHE}}$ in (5) is zero. Therefore, similar to STT, SOT-based i-MTJ suffers from long delays due to the incubation period. This makes SOT-based i-MTJs and STT-MRAMs not proper for high-speed memory application (<1ns) due to the significant effect of the incubation delay. For SOT-based p-MTJ, the incubation delay will be very negligible. In this scenario, $\overrightarrow{\sigma_{SHE}}$ is orthogonal to $\overrightarrow{m_{FL}}$, hence their multiplication is non-zero that generates a large SOT, which enables an ultra-fast switching process at the cost of making the switching process stochastic [27]. The big SOT induced by $\overrightarrow{m_{FL}} \times \overrightarrow{\sigma_{SHE}}$ rotates $\overrightarrow{m_{FL}}$ very fast and makes it collinear with the $\overrightarrow{\sigma_{SHE}}$. As long as $I_{SOT}$ exists, $\overrightarrow{m_{FL}}$ stays in this position. By removing $I_{SOT}$, $\overrightarrow{m_{FL}}$ tends to relax in one of the perpendicular anisotropy axes. Choosing between these two steady-state directions (e.g. $+z$ or $-z$) is decided by *thermal fluctuation*. Therefore, deterministic switching is not possible using a single $I_{SOT}$, and an external field is needed.

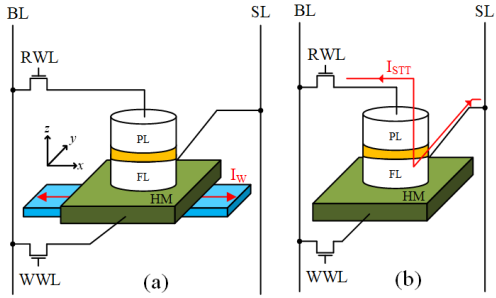

Fig. 4. (a) p-MTJ SOT-MRAM cell deterministically switched by $H_{ext}$. (b) p-MTJ SOT-MRAM cell deterministically switched by $I_{STT}$.

According to (5), two external torques other than SOT can be used to enable deterministic switching; $H_{ext}$ and $I_{STT}$. Conventionally, $H_{ext}$ is used for deterministic switching as shown in Fig. 4(a) [39]. In this figure, the uniaxial perpendicular magnetic anisotropy axis of the MTJ is along the $z$ direction and $I_{SOT}$ flows in the $y$ direction generating $\vec{\sigma}_{SHE}$ along the $x$-axis. A $H_{ext}$ is applied to the MTJ by putting a conductive line underneath the FL of the MTJ, e.g. by passing a bidirectional current ($I_W$) through the line. When $H_{ext}$ is in the $+y$ direction and $I_{SOT}$ flows from BL to SL, the MTJ can be switched to AP-state, while flowing $I_{SOT}$ from SL to BL enables switching to P-state. The reverse scenario can be concluded for $H_{ext}$ in $-y$ direction. However, in scaled technology nodes, adding a permanent magnetic field degrades thermal stability, scalability, and increases the complexity.

To this end, instead of $H_{ext}$, $I_{STT}$ may be used enabling deterministic switching (Fig. 4(b)) [40]. To do this, $M_{read}$ creates $I_{STT}$ during write operation; hence no extra transistor is added to the cell. As experimentally demonstrated in [27], if $J_{SOT}$ is bigger than $J_{c.SOT}$, the switching process is governed by the SOT and is almost independent of the amplitude of $I_{STT}$.

$J_{c.SOT}$ is defined as the critical current in which $\overrightarrow{m_{FL}}$ rotates to the in-plane orientation. It means the size of $M_{read}$ can still be optimized based on read operation requirements.

Either we use $H_{ext}$ or $I_{STT}$, to enable switching process, both $I_{SOT}$ and $H_{ext}/I_{STT}$ are active. The spin torque induced by $I_{SOT}$ rapidly rotates $\overrightarrow{m_{FL}}$ from $z$ direction to $x$ direction. However, due to the torque induced by $H_{ext}/I_{STT}$, $\overrightarrow{m_{FL}}$ passes in-plane direction and is stabilized between $x$-axis and $z$-axis. Switching to AP-state and P-state are shown in Fig. 5 in which $m_x$, $m_y$ and $m_z$ define $\overrightarrow{m_{FL}}$ elements along $x$-, $y$- and $z$-axis, respectively. According to this figure, $\overrightarrow{m_{FL}}$ will not be switched to the easy axis as long as the $I_{SOT}$ is active. Therefore, to accomplish switching, $I_{SOT}$ must be removed at an appropriate time while $H_{ext}/I_{STT}$ must be kept active until the switching is completed.

### C. Multilevel SOT-MRAM

As mentioned before, the main disadvantage of the conventional SOT-MRAM cell is their area overhead in comparison with other MRAM cells. Therefore, such SOT-MRAMs are not appropriate for high-density memory applications [36]. Although using a Schottky diode instead of a transistor may help, still more practical solutions are needed.

A known technique is multilevel SOT-MRAM cells. In such cells, multiple bits are stored in each memory cell with more MTJ elements in one cell. In SOT-MRAMs, multilevel design concepts are studied in [41], [42]. In [41], two series MTJ multi-level cell (S-MLC) and Parallel MTJ multi-level cell (P-MLC) shown in Fig. 6(a) and (b) have been proposed. In S-MLC, two series MTJs construct the storage elements of the cell in which only the FL of MTJ$_2$ is in contact with the HM. Therefore, in this structure, MTJ$_1$ and MTJ$_2$ can be programmed by STT and SOT mechanisms, respectively. However, both MTJs cannot be programmed at the same time; MTJ$_1$ is written first and then MTJ$_2$ is written. This approach eliminates write–disturb failure in the cell; even if an unwanted switching occurs in MTJ$_2$ during programming of MTJ$_1$, the correct data can be written back to MTJ$_2$ in the next cycle. However, during the programming of MTJ$_1$, large write current flows through the MTJ$_2$ as well degrading the reliability of MTJ$_2$. To deal with this issue, in P-MLC, the FLs of both MTJs are in contact with the HM and the MTJs are connected in parallel to each other.

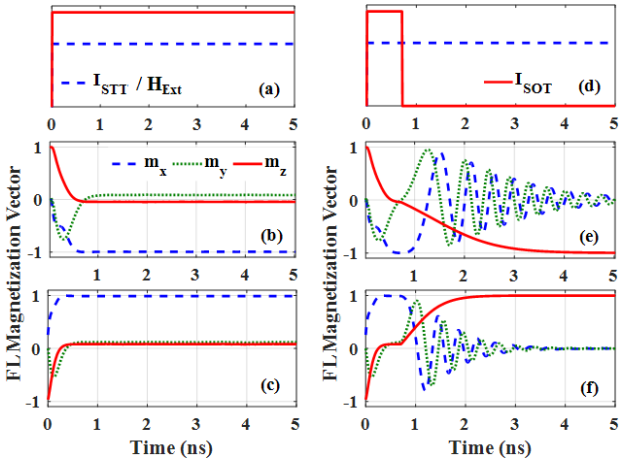

Fig. 5. (a) If $I_{SOT}$ is not removed during switching, FL magnetization is stabilized at a specific orientation between in-plane and $z$-axis during switching to (b) AP-state and (c) P-state. (d) If $I_{SOT}$ is removed during switching, FL magnetization is aligned with the easy axis during switching to (e) AP-state and (f) P-state.

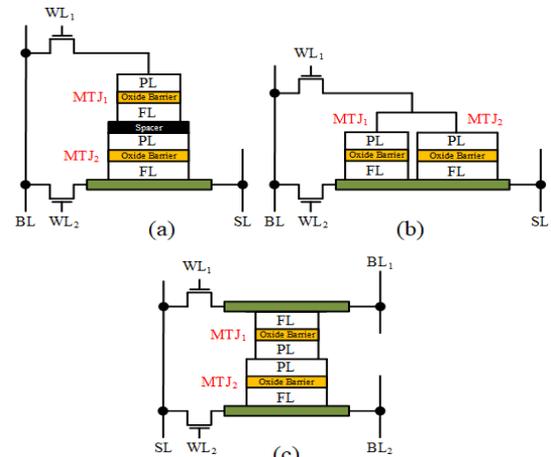

Fig. 6. (a) S-MLC, (b) P-MLC, (c) Multilevel SOT-MRAM cells with separated HMs.

Therefore, both MTJs are programmed by the SOT mechanism. On the other hand, both MTJs are in the same SOT current path, hence they must have different $J_c$'s. One approach to achieve such a difference is to increase cross-sectional area of the HM underneath MTJ$_2$. This reduces the spin injection efficiency of SOT mechanism in MTJ$_2$; therefore, this increases $J_c$ of MTJ$_2$ making it bigger than that of MTJ$_1$. To program the memory cell, the MTJ with a larger $J_c$ (i.e. MTJ$_2$) is written first and then MTJ$_1$ is programmed by passing a $I_{SOT}$ lower than the critical current of MTJ$_2$. Similar to the P-MLC, another multilevel cell in which the write operation is accomplished by the SOT mechanism has been proposed in [42]. In this cell, shown in Fig. 6(c), the storage element is built by stacking two MTJs with a shared PL. In this cell, FLs of MTJ$_1$ and MTJ$_2$ are in contact with separate non-magnetic HMs enabling separate SOT-based switching for each MTJ. During write operation of MTJ$_1$, I$_{SOT}$ flows through the top HM by asserting WL$_1$ signal while WL$_2$ signal is switched off. When MTJ$_1$ is programmed, the sneak current passing through the stack MTJ can be eliminated by floating BL$_2$. On the other hand, switching of MTJ$_2$ is enabled by asserting WL$_2$ signal while WL$_1$ signal is off and BL$_1$ is floating. The MTJs may have the same critical currents making this structure more power-efficient in comparison with the S-MLC and the P-MLC. To read out the stored data in multilevel cells, four resistance states are needed. To meet such requirement, MTJs are implemented using different cross-sectional areas; MTJ$_1$ is designed with the minimum feature size of the technology while MTJ$_2$ is chosen larger. Using different sizes for MTJs leads to different thermal stabilities, which is challenging during read.

### D. Thermally Assisted SOT

One promising method that recently attracted attentions is to heat the MTJ in order to decrease the energy and the delay of the switching [43]. In fact, heating the MTJ decreases the saturation magnetization ($M_S$) of the free layer, which leads to lower uniaxial anisotropy field. Hence, deviating the magnetic direction of the free layer from easy axis to either switch it or starting the oscillation will be easier. One way to heat the FL is to use Joule-heating around the tunnel barrier by passing a current through the MTJ to ease switching - called thermally assisted switching (TAS) [43]. In this method, a modified type of MTJ is used such that the FL consists of two layers including a ferromagnetic layer pinned with a low blocking temperature (T$_B$<160ºC) antiferromagnet (e.g. FeMn). Hence, at room temperature, the FL is very stable. However, the temperature of the MTJ will be increased to a temperature higher than T$_B$ during switching. Hence, a small external magnetic field or a small STT current can switch the MTJ (Fig. 7).

Although TAS method decreases the required energy consumption for switching, it comes with some limitations and disadvantages. First, a modified MTJ is needed, and it is not efficient for typical MTJs. Second, as TAS uses a current passing through MTJ for heating, it causes reliability issues.

There is a possibility of switching the MTJ while heating it up that is a failure in neuromorphic computing application where elevating the temperature is just used to assist the MTJ-based neurons to switch and it is not supposed to switch them. Moreover, the minimum time needed to heat up the MTJ is 500ps in TAS method which put a limit on maximum achievable speed. Finally, it is energy efficiency is lower than laser assisted method as will be discussed in the next paragraph.

### E. Optically Assisted spintronic Memories

With ever-growing photonic integrated circuits (PIC) with a higher level of integration and the promising vision of scaling down the photonic components on a chip [44], several projects with high investment from industry and academia around the world are exploring the possibility of bringing photonics as close as possible to electronics. Applications for such technology include wired (e.g. through waveguides on chip or optical fibers) and wireless high-speed high bandwidth communications, sensing application, ADC etc.

- *Optical STT-MRAM:*

Among different research projects, SPICE, a European-funded project, has been exploring the possibility of using the known all-optical switching (AOS) phenomena in different materials, and bringing it to a higher level with a PIC combined with a spintronic and an electronic integrated circuits to develop the first ever fully integrated optical MRAM with the lowest energy figure at the highest speed [45]. This co-integration uses AOS as a writing mechanism for optically accessible MTJs in order to harvest the high speed and low energy density of AOS [46]-[48]. Experiments show two particular cases of AOS: helicity dependent switching with a circularly polarized light pulse (HDS) and helicity independent switching (HIS) with an unpolarized light pulse [49]. HDS occurs in a small energy window around the AOS switching threshold but is able to fully control the final magnetization orientation of the magnetic layer by controlling the circular polarization direction of the input light. This direct control can potentially lead to an all optical MRAM memory which could either be envisioned as a long-distance high-speed memory in a photonic electronic co-integrated scheme or as a memory element in an all optical computer. However, using HDS requires a very precise control of the light energy, which would likely be a large strain on the fabrication processes to obtain sufficiently reliable devices. HIS benefits from a larger energy window but is a pure toggle phenomenon, *i.e.* the FL magnetization is reversed after every light pulse above the AOS threshold. This forces an extra control of the incoming light pulses and the eventual memory. As such, in a HIS writing scheme, reading the memory before the writing light pulse is sent becomes a necessity. This slows down the writing process, makes the electrical side of the integration complex, and increases the energy budget of the memory. Another challenge of photonic-electronic integration is the footprint mismatch between photonic ($\mu m^2$) and electronic ($nm^2$). The integration density of such photonic spintronic electronic memory could be very low compared to electrical equivalent if this mismatch is not reduced either with more compact photonic devices or with other integration schemes for AOS *e.g.* plasmonic device.

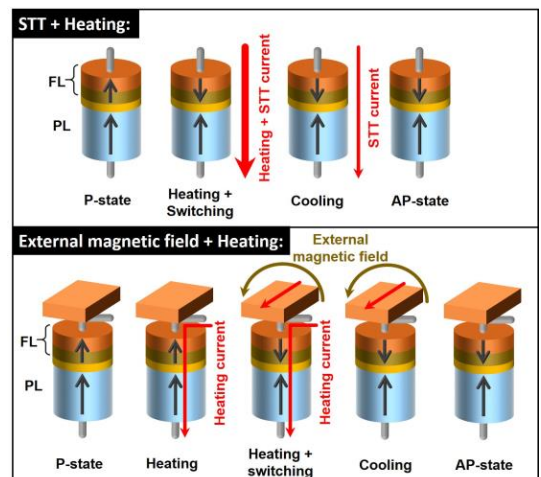

Fig. 7. Thermally assisted MRAM. The heating is combined with STT current or external magnetic field in order to switch the MTJ

- *Optically assisted SOT-MRAM:*

Repeatedly heating an MTJ free layer could degrade the device reliability and/or its lifetime due to the repeated thermal stress brought by the optical switching of the memory. To solve these issues, we propose the use of SOT-MRAM instead, which removes the need of directly hitting the MTJs; instead, the heavy-metal layer, which may lead to a higher reliability with a possibility of cooling down of the raised temperature at higher speed. This approach is expected to improve the reliability of the MTJ devices while a more proper approach can be used to cool down the HM and so the MTJ faster.

Another envisioned advantage of the use of such approach is the possibility of taking advantage of Seebeck effect [50] leading to a faster switching by assuming the exposure of the heavy metal to a femto (pico)-second laser pulse. However, this requires proper material stacks and engineering not to allow the laser pulse hitting the MTJ stack. This can be a research path for years which benefits from the advantage of the speed of photonics and scalability of spintronics as well as low energy to implement high-speed and low-power memories.

### F. Strain-Assisted SOT switching

One of the emerging fields within spintronics is assisting switching magnetization of the magnetic layers by the use of strain through magnetoelectric-coupling. Such technology for memory applications has shown high potential for changing the magnetic properties of ferromagnetic materials with a low energy [51]. To achieve a low energy for switching, a ferroelectric layer with high electromechanical coupling and piezoelectric coefficient, e.g. [Pb(Mn1/3Nb2/3) O3] (1-x)-[PbTiO3]x (PMN-PT), can be used [52]. Some works show the use of other piezoelectric materials such as Lead-Zirconate-Titanate (PZT) for exerting a magnetic field on a magnetization layer for switching. Taking such an approach removes the requirement of an external magnetic field for initiating the switching completed by SOT using Perpendicular Magnetic Anisotropy layers. Some works propose the use of acoustic waves by applying a femtosecond laser pulse, which creates a surface acoustic waves (SAW). By using this approach, high speed and low energy memory switching is achievable without taking the energy consumption of photonic layer for distributing the laser pulse on the target cells. In fact, this approach is similar to the photonic-assisted switching while strain of the piezoelectric layer is used to assist the switching rather than the heating. For Strain-assisted SOT memory, the piezoelectric layer is biased with the required voltage depending on the input data. This can help the magnetization switching of the magnetic FL. Such approach can be used for many other applications of SOT e.g. computing systems. The main disadvantage of such technology is the complexity of the design as well as their scalability. Although several groups have shown some successful experiments for switching the magnetization of magnetic material it is still far from reality as the size of the devices used are much larger (micrometer size) than the state-of-the-art spintronic cell sizes (10nm diameter and few nm thickness), and scaling the magnetic device and so the piezoelectric layer to target individual cells, will scale the amount of magnetostriction of piezoelectric layers (e.g. Ni), which is not enough to change the magnetization of a magnetic layer or even help enough (less than 10%). Nevertheless, other materials have been explored to improve this or a larger piezoelectric layer can be stressed under an array of cells for a better efficiency and then writing electrically.

### III. SOT-MRAM VERSUS OTHER EMERGING TECHNOLOGIES:

One of the main purposes of proposing novel memories was to realize a *universal memory* that can be used for different purposes ranging from caches to hard-disks; very challenging as each memory offer different features. The main emerging memory technologies include Phase-Change Random-Access-Memory (PCRAM), Resistive RAM (ReRAM), Voltage-Controlled Magnetic Anisotropy MRAM (VCMA-MRAM(MeRAM), STT-MRAM and SOT-MRAM. Almost all these memories suffer from high write currents and high write latencies. Spin-based memories have shown better endurance in comparison to the ReRAMs and PCRAMs. PCRAMs suffer from a much larger write latency compared to other emerging technologies. The retention of these memories is similar and all are non-volatile.

Overall, spin memories are more promising than others while still suffer from high write energy and long latency, which makes them not fitted well for high-speed caches. Table 1 summarizes comparison of spin-based memories.

Table1. Spin-based memory comparison

| Metrics | In-plan STT-MRAM (1T1MTJ) | VCMA-MRAM (MeRAM) | SOT-MRAM |
|---|---|---|---|
| Write(Read) Energy/bit | 22,2pJ(13,8pJ) [53], 0.7-3pJ [59] | 0.27nJ (0.21nJ) [56] | 0.334nJ (0.254nJ) [55] 1-2pJ [57] 0.4-0.5nJ[58], 0.3-3pJ [59] |
| Write Latency (ns) | 11 [54] 17,2 [57] 30 [55] 40,63 [56], 0.5-20 [59] | 9,4 [56] | 1,36 [54], 1 [58] 0.3-5 [59, smaller for SOT] |
| Read Latency (ns) | 1,2 [54] 11,9 [57] 30 [55] 10,63 [53] | 5 [56] | 1,13 [54] |

To solve this issue, we anticipate any of the following paths:
1. *Other switching mechanisms than torque*: Such example is AOS that can be used to achieve significantly higher speed enabled by femtosecond laser pulses through Seebeck effect, thermal effect or any other untapped effects due to the interaction of spin-based devices with laser. This approach is still in nascent stage and SPICE project is a unique project taking this approach to the next step to develop a proof-of-concept AOS memory. Although the footprint of the photonic components are orders of magnitudes larger than spintronics or electronics components footprints.

2. *Assisting techniques*: Such techniques can be used for assisting the switching process with the cost of more complexity. Among others, photonic/optically-assisted, thermally-assisted, strain-assisted are some examples showing some promises for future fast and low-power memories. These techniques are mostly used to ease the switching of the magnetization. One can assess the question over the improvement that can be achieved by such techniques.

3. *Novel SOT-MRAMs*: One of the main challenge with STT-MRAMs or in-plane SOT-MRAMs is the large incubation delay that can be improved by using novel materials or novel architectures for removing the incubation delay of the switching mechanism [60]. Due to the nascent stage of developing such memories, there are still high potentials for novel materials to improve the spin efficiency of such technologies to ease the switching with lower powers and higher speeds.

## IV. SOT-BASED OSCILLATORS

Nowadays, traditional on-chip oscillators face several challenges in terms of integration due to CMOS scaling and power consumption. To deal with these obstacles, spin hall nano-oscillator (SHNO) and spin torque nano-oscillator (STNO) can be considered as appealing candidates as new devices beyond CMOS for radio frequency (RF) applications. Each of the mechanisms has its advantages, but in the following we will focus on the different types of (SHNOs). The SHNOs can be divided in two-terminal and three-terminal devices.

### A. Two terminal SHNOs

The two terminal SHNOs are based on the anisotropic magnetoresistance (AMR) effect as readout mechanism. There are three different two terminal SHNOs developed in the literature. Nano-gap [61], nano-constriction [62] and nanowire [63] shown in Fig. 8 are examples of such SHNOs. In spin Hall effect (SHE) driven SHNOs, spin current is produced by placing a heavy metal (HM) and pass the dc current along it. Due to high spin-orbit interaction in the HM, the spins with opposite direction accumulate at opposite metal interfaces. If a ferromagnetic (FM) layer is placed in contact with the HM, the accumulated spins can diffuse into FM and exert a torque on the magnetization of the FM to overcome the damping torque in

LLGS equation and achieve steady state precession of magnetization. In practice, a high spin current density is required, so SHNO devices are fabricated in nanoscale geometries to substantially increase the current and consequently, the spin density. The fabrication process of such devices is much easier than that of the STNO because of their planar geometry. Moreover, in two terminal SHNOs, the magnetization dynamics can be studied in frequency, time and phase domains by optical means [64-66]. SHNOs exhibit very exciting phenomena such as broad frequency tunability [67], producing localized and propagating spin-waves for magnonic devices [68, 69], lower oscillation threshold current, and higher signal coherency compared to the STNOs. However, the output power of SHNOs is lower, because they work based on AMR effect. Furthermore, the coherency of their output signal which is determined by their linewidth is high compared to their semiconductor rivals. Hence, some solutions are required to improve the output power of the SHNOs. Synchronization of multiple oscillators is an efficient approach by which the oscillating magnetic region expands its volume by providing more applied dc current [70]. Once SHNOs are placed in proximity, their precessing volumes overlap and the magnetic exchange and dipolar interactions force the oscillators to operate in unison with a finite phase difference. The SHNOs in this state are called to be phase-locked as the coupling between each pair of them is defined by their phase difference [71].

Synchronization of SHNOs was first shown in a work done by Awad *et al.* [64]. In their work, robust synchronization of up to nine SHNOs fabricated in a chain was achieved in which upon synchronization, the signal linewidth substantially improved proportional to the number of synchronized oscillators ($N$) and the peak power increased as ($N^2$). Later, two—dimensional (2D) synchronization of SHNOs was demonstrated by the same group [65] where a massive array of 64 SHNOs placed as an 8×8 SHNO array synchronized as an ensemble. The synchronized array operating at 10 GH outputted an extremely coherent signal with 59 kHz linewidth reaching a record high quality factor of Q=170,000 for spin-based oscillators. 2D SHNO arrays offer a very feasible path towards large scale coupled oscillatory networks as they can be easily scaled down to 20 nm [72]. There are many non-conventional computing paradigms based on 2D coupled oscillatory networks, including, but not limited to, vertex coloring of the graphs which is a class of combinatorial problem [73], pattern matching [74], auto-associative memory computing [75, 76], image segmentation [77], and Ising machine for solving combinatorial optimization problems. What all these paradigms have in common is that all the inputs and memorized patterns and information are encoded in the coupling between the oscillators defined by their phase differences. So, it is important to have access to the individual oscillators at nanoscale to tune the coupling for training purposes.

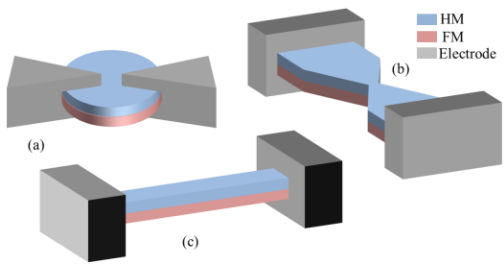

Fig. 8. (a) nano-gap, (b) nano-constriction, and (c) nanowire SHNO.

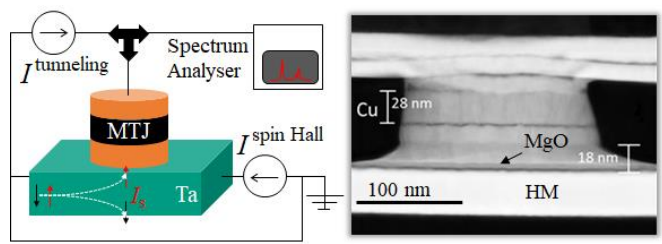

Fig. 9. Schematic of the microwave emission measurement circuit setup for spin-polarized current-induced nano-oscillator devices on the left. On the right, a cross sectional SEM image prepared with focused ion beam.

### B. Three-terminal SHNOs

Despite the attractiveness of SHNOs, their output powers are far below the output powers of STNOs. In an effort to improve the output power of the oscillator, a three terminal approach was developed [78], in which the free layer (FL) of an MTJ is excited by a pure spin current generated by the spin orbit torque (SOT) in a HM layer below the MTJ (Fig. 9). Such device can be also understood as a conventional two terminal SHNO, which employs the tunneling magnetoresistance (TMR) effect instead of the AMR effect to increase the efficiency of the readout mechanism. Liu *et al.* [78] could show that using the spin current injected by the SOT created oscillations of 0.25 nW at a relatively lower tunneling current of 60 μA. Later, Tarequzzaman *et al.* [79] showed that in devices with higher TMR effects output powers of above 8 nW can be easily reached with lower tunneling currents. Despite the improvement, this output power is still very far from the output power than can be achieved with conventional STNOs; hundreds of nW for single devices [80] and has been shown to reach 14 μW with 8 synchronized vortex devices [81].

However, in this type of devices the output power scales with the square of the tunneling current. Therefore, increasing the tunneling current offers the possibility of increasing the output power significantly. However, at some point, both spin injection mechanisms, the SOT and the spin transfer torque (STT), contribute to the excitation of the FL. Using this combined spin injection, the same device can reach a total output power of 48 nW. The system response can be coherently quantified as a function of the total injected spin current density with a critical spin current density of $4.52 \pm 0.18 \times 10^9 h/2$ e$^{-1}$ Am$^{-2}$. This is equivalent to a critical current density of 5 nAm$^{-2}$ or a critical SOT current density of 150 nAm$^{-2}$. This level of spin injection is far more than it could be possible to achieve with STT alone without damaging the tunnel barrier or with SOT alone without destroying the spin Hall line. It is important to note that one of the original motivations to introduce 3-terminal devices was the possibility of exciting large orbit dynamics in the FL without disturbing the delicate tunnel barrier. All things being equal it should be possible to replace STT current by SOT current in such a system in order to achieve the same current injection, according to:

$$J^{spin} = (\eta J^{tunneling} + \theta J^{SpinHall}) h/2e^{-1} \qquad (1)$$

This relation shows that the tunneling spin injection efficiency η and the spin Hall angle θ are key parameters determining how efficiently a spin tunneling current can be replaced by a spin Hall current to excite dynamics. With MgO tunnel barriers η can easily above 50%, while the largest spin Hall angles are only 20%. Although, the spin Hall current does not flow through the tunnel barrier, the current densities in the

HM line have practical limitations. Due to the scattering-based spin Hall effect, the HM materials are highly resistive, which results in a large heat dissipation close to the tunnel junction. This limits the spin Hall current to a practical maximum as high temperatures are detrimental for the MTJ and electromigration effects cause irreversible damage to the HM line.

Tarequzzaman *et al.* reported that in their Ta line dissipates 80 % of the output power, meaning that they expect a 5-fold increase of the output power by removing the effect of the HM resistance alone [79]. The goal should be to maximize the spin Hall current densities below the MTJ, while minimizing the total resistance of the HM line. This can be achieved by the geometries used in nano-constriction SHNO and further improved by depositing low resistivity material on the HM leads to the contact pads. In addition, increasing the oscillator resistance through reducing the size and increasing the R×A of the MTJ nanopillar can improve the power transfer. For typical STNOs increasing the R×A is not a viable strategy as it decreases the maximum excitation current and the output power [80]. However, for three terminal SHNOs the excitation can be always achieved through the SOT. Thus, the three terminal SHNOs can make full use of separating the excitation and sensing mechanisms. First results as a function of the R×A and a bias current of 50 μA showed a maximum output power of 12.5 nW at a R×A of 34 Ωμm$^2$ [82].

Besides these experimental efforts on three-terminal SHNOs, there have been some efforts on macrospin modelling of these structures [83], which was able to model all relevant characteristics of the oscillators. Another theoretical work on micromagnetic models included a scalable synchronization scheme for SHNO MTJs in parallel [84]. As reviewed in [85], Pt and Ta are appropriate materials to generate large spin-orbit couplings. To provide enough pure spin-current induced by SHE, different FM materials including NiFe, $Co_{40}Fe_{40}B_{20}$, Co/Ni multilayers with perpendicular magnetic anisotropy and the ferrimagnetic insulator YIG have been applied [86].

## V. SPIN-BASED COMPUTING SYSTEMS

Von-Neumann computing will inevitably within the next 10 years reach a fundamental limit because of technical reasons [87] such as limitations on manufacturing, interconnects (speed and density), transistor technology (the width of the gate dielectric is limited by atomic spacing limit), power constraints (power scaling is slower than technology scaling) etc. [88]. Brain-inspired computing, i.e. mimicking the behavior of the brain and implementing it into hardware, especially for cognitive computing has shown great potentials proven theoretically and experimentally. IBM's TrueNorth and Intel's Loihi are examples of neuromorphic computing systems (NCSs) [89-91].

NCSs use many parallel processors (neurons) communicating using simple messages (spikes) or continuous interactions (oscillations) mediated by programmable memory units (synapses) as well as routing circuitries (Fig 10 (a)). CMOS implementation of NCSs is area- and power-inefficient [92]. Such inefficiencies have driven a significant effort to explore implementation of beyond-CMOS NCSs, where synapses have been implemented using memristors [93, 94] as well as spin-based materials [95-97], and neurons are implemented using STT-MTJ, mutually synchronized SOT-MTJ [93], and phase change materials (PCMs) [98]. In the following subsections, we review the spintronic NCSs with a focus on the future vision of the SOT-based NCSs and the entailed challenges.

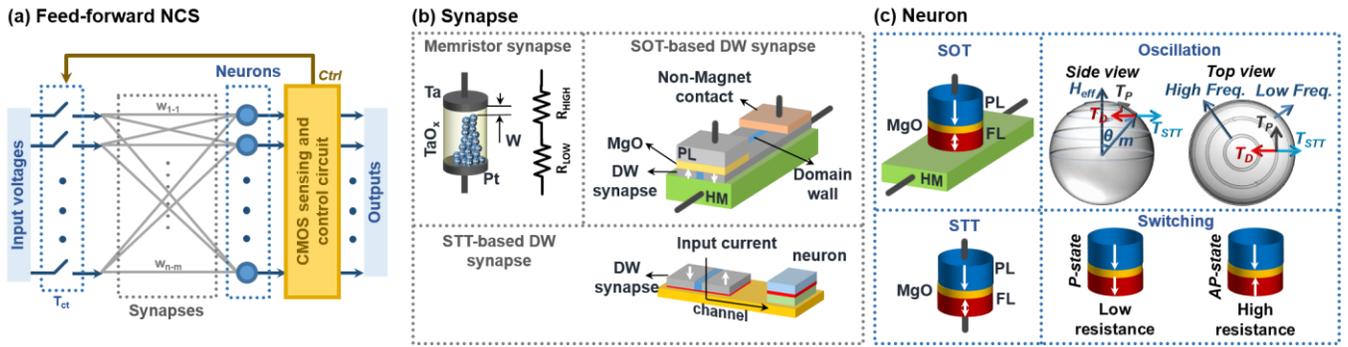

Fig. 10. (a) The block diagram of the feed-forward NCS. Different spin-based implementation of (b) synapse and (c) neuron.

### A. Oscillation versus switching-based NCS

In spin-based NCSs, the magnetic reversal and the magnetic oscillation of FL magnetic moment in STT/SOT-MTJs can be used to mimic neuron firing. However, each of these two approaches has pros and cons. The main advantage of utilizing oscillation compared with magnetic reversal is the considerably lower power consumption. This is due to the fact that, first, starting the oscillation of magnetic moment in a device needs a lower power consumption compared with the magnetic reversal (critical current: $\sim 10^6 Acm^{-2}$ [80] vs $\sim 10^{-7} Acm^{-2}$ [99]). Second, in contrast with switching-based NCS that needs to consume power in order to switch back the magnetic moment before the next processing round (called recovery phase), in the oscillation-based NCSs, there is no need for recovery phase.

On the other hand, sensing approach of the oscillation-based NCS is significantly more complicated and more power consuming than the switching-based NCSs. In the switching-based NCS, the sensing circuit should distinguish between AP- and P-state resistances. Considering the large difference between these two resistances (TMR>100%), the sensing circuit can be designed with low power consumption using a comparator [100, 101]. However, in oscillation-based NCS, the sensing circuit needs to sense the oscillating signal with a nano-Watt output power. This means the output signal across the nano-oscillator, first, should be amplified with a low noise amplifier [102], and then, converted to a DC signal detected by a power detection (PD) circuit. Finally, the DC output of PD is compared with a reference voltage that leads to a large-area and high-power especially at higher frequencies. One approach is to use switching elements themselves as PDs. This can be done thanks to the spin diode effect, which makes use of a resonant excitation of the dynamic modes of a confined magnetic nanostructure. When such a structure is excited with an RF current close to the natural frequencies within the system, the convolution between the RF current and the time changing resistance of the structure results in a DC output if the frequency of the input signal is within a certain range of the natural modes.

In vortex nano-structures, this effect is used to completely expel a vortex, which results in a very large DC output [103]. Recent experiments show that besides this digital detection scheme, it is possible to obtain a DC output which is proportional to the frequency of the input signal [104]. Using MTJ elements both as frequency generators and frequency detectors within the same NCS has a potential to enable new network architectures.

### B. Spin-based neuron and synapse implementation

In spin-based NCs, both switching [102, 105] and oscillation [106] of FL in the STT-MTJ and SOT-MTJ can be used to mimic neuron's functionality while crossbar array of memristors [93, 94], SOT-based memory [107], and STT-based [95] and SOT-based [96, 97], [108] domain wall MTJ (DW-MTJ) can be used to mimic synaptic functionality. In case of utilizing memristor as synapse (Fig. 10(b)), the amount of injected pure current to the neuron can be tuned by setting the resistance of the memristors using an electric signal flowing through them. As a result, the amount of pure current passing through the neuron is tunable. The general operation of the SOT-based memory is very similar to the memristor. The resistance of the SOT-based memory can be tuned by passing electrical current through it. Hence, the SOT device can be used as analog memory (synapse) to store the synaptic weights. On the other hand, DW-MTJ (Fig. 10(b)) can acts as synapse by controlling the amount of polarized current passing through the spin-based neuron.

*SOT-based neuron:* SOT-based devices are used to mimic different types of neurons in the literature. There are several SOT-based implementation of neuron's threshold transfer function by simple SOT-MTJ [105] and mixture of domain wall and SOT-MTJ [109]. SOT-MTJs are implemented using MTJ with a heavy metal (HM) layer adjacent to the FL of MTJ. In [110] and [111], the stochastic nature of SOT-MTJ switching is used to mimic the probabilistic spiking nature of neurons. Moreover, it is shown in [112] that the STT/SOT-MTJs can be used to generate output signal consist of a single spike or a discrete group of spikes (bursting) at frequencies of tens of GHz, a breakthrough for the future neuromorphic signal processing applications. Finally, SOT-MTJs have been utilized in other forms of computation methods (like Ising computation) where the conventional computing is energy inefficient due to huge amount of input data [113].

*STT-based neuron:* STT-based implementation of different artificial neurons have been extensively investigated. An STT-

based NCS for vowel recognition has been proposed in [106]. It requires conversion of sound's frequency to a much higher frequency which compromises the power advantage of using oscillators. Moreover, STT devices based on domain wall motion magnetic strip has shown energy efficient implementation of a neuron with non-linear analog output [114], [115]. In [116], an STT-MTJ neuron is proposed which is able to mimic the integrate-and-fire behavior of the biological neurons. Note that using SOT decreases the stress on the MTJ device, which may lead to a higher reliability of the device. The same approach applies to the use of SOT-MTJ oscillators (3-terminals) where the current passing through the HM leads to an oscillation. Use of other types of mutually synchronized SOT-MTJ oscillators have been proposed in literature too [117].

### C. Optically Assisted Spin-based NCSs

Given the possibility of high level of integration of photonic components, we have been leading activities in developing application including memories and NCSs benefiting from high-speed photonics, matureness of electronics and small footprint of spintronics. In this part, we explore the achievements within this field and will sketch the future of this technology for interested researchers.

In laser assisted (LA) method, the laser illumination will be used to heat up the STT/SOT-MTJs in order to ease the switching/oscillation of FL (Fig. 11). The crossbar array of memristors with tunable resistance is used as synapses and STT/SOT-MTJs are used to mimic neuron activity. The temperature of SOT-MTJ (STT-MTJ) will be elevated by illuminating a narrow laser pulse on its HM (top contact) in order to ease switching/oscillation of it [102], [118]. The general operation of the LA-NCS is as follows: First, in a calibration phase, the temperature of STT/SOT-MTJs will be increased to 100°C and stabilized. After calibration, the processing phase will be started that can be divided into two steps including stimulation and recovery steps [102]. Note that the processing phase will be done at elevated temperature (around 100°C) which leads to lower energy consumption and faster operation of LA-NCS.

In stimulation step, the input voltages will be applied to the inputs of crossbar array and based on the inputs, one of the STT/SOT-MTJ neurons (already set in AP-state) will switch/oscillate that will be detected by a sensing circuit translated to firing. The sensing circuit should use track and terminate method [100, 102], [106] in order to minimize the energy consumption of the NCS. Although the photonic layer increases the energy consumption of the NCS, the overall energy improvement (~86%) and speed (~84%) of the LA-NCS due to working at elevated temperatures is significantly higher. Immediately after detecting switching/oscillation by sensing circuit, the recovery phase will start in which (a) the NCS will be prepared for the next stimulation, (b) the corresponding input in the post-layer will be activated and (c) the temperature reduction during the stimulation phase will be compensated by illuminating a short laser pulse. Note that for oscillation-based

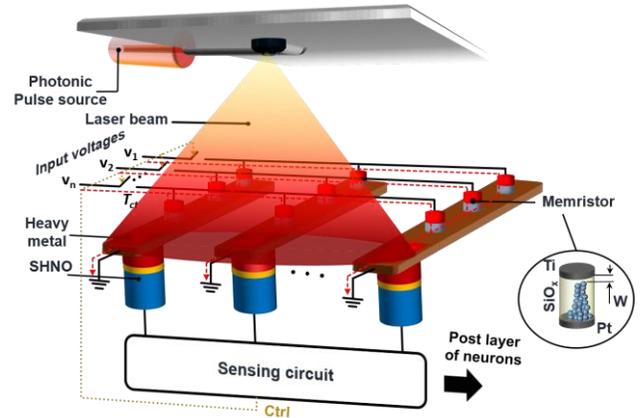

Fig. 11. The laser-assisted SOT-based NCS. The laser pulse is illuminated on the heavy metal layer of the SOT-MTJ to heat it up.

NCS, switching back the neurons is not needed (no need for task (a)).

Both switching approach using a single STT/SOT-MTJ (Fig. 11) with possibility of switching and oscillation using an array of SOT-MTJs with possibility of mutual synchronization (to increase the output power) can be taken. As mentioned before, passing current through HM will lead to an SOT acting on the magnetic direction of the FL magnetization. In case, the HM current will be high enough, the SOT will be able to start oscillation or switch the FL magnetic direction. Note that passing current through SOT's HM instead of the MTJ offers two important advantages: 1. higher reliability of the SOT-MTJ due to less stress on the tunneling barrier, 2. Lower charge current for SOT-MTJ by increasing the MTJ cross-section area compared with the HM cross-section area. The central challenge of SOT-MTJ is its low efficiency of the charge-to-spin current conversion, quantified by the spin Hall angle $\theta$ which relates the spin current generated in the HM with the charge current flowing on it by $J_s^{HM} = (\theta J_c^{SpinHall})\hbar/e$. The reported value of $\theta$ is in the range of $2 - 23\%$, which means, in the best-case scenario, only 23% of the charge current through the HM is converted to spin current. This leads to lower energy efficiency of the SOT-MTJ [71]. Hence, increasing $\theta$ can effectively reduce the energy consumption and delay of the SOT-based NCS. These can be done through material engineering of SOT-MTJ and/or heating up them. Recently, a spin-Hall angle of 53% with W on CMOS compatible substrate was reported [80] that proves the possibility of improving the efficiency of the SOT-MTJ through material engineering.

### D. 1.2.2 Strain-assisted NCS

As mentioned, straintronics or using the stress exposed on a piezoelectric material is a new field of research and it has been nicely studied in some literature [119] for different applications mainly for memory [120-123], as well as micro-nano motors, micro fluidics, and microwave generators. In literature, straintronics, very recently, has been used to develop novel logics and a novel neuron structure by [124] with the possibility of being integrated with a memristor crossbar array for synapse implementation leading to a potential neuromorphic computing system. In strain-assisted SOT-based neuromorphic computing system, a similar approach can be taken by replacing the neurons with strain assisted SOT and the synaptic weights can be implemented using memristors. The advantage of using SOT is that the fabrication process is not complicated while no stress is put on the MTJ improving the reliability of the design. Furthermore, the piezoelectric biasing can be applied as a biasing for neurons in the classification layer of the proposed neuromorphic computing system.

### E. SOT-Based Approximate Computing System

Approximate Computing, which is based on inexact logic for the least significant bits (LSBs) of any binary operation, can be at different levels changing from circuit simplification, new devices and novel architectures. However, almost all of the works have been done so far within spintronics and non-volatile memories such as [125]-[126] are focusing the possibility of using less precise data storage for storing non-important data to lower the total power of the system. This is widely applicable to image processing and so applicable to implementation of artificial neural networks.

Within approximate computing, the works we have done on STT-MRAMs [127] can be applied to the SOR-MRAMs too. To apply this, we make less accurate SOT-MRAM cell in a similar manner of the STT-MRAM for LSBs and more accurate ones for MSBs. For SOT-MRAMs, however, we may have more freedom as we can play with the heavy-metal area too.

## VI. OTHER APPLICATIONS

In previous sections, we investigated the possibility of using SOT technology in computing and storage systems. Such systems are usually employing binary data during computation. However, when these systems need to communicate with analog world, some analog/mixed-signal circuits are required. The straightforward approach is to design analog/mixed-signal circuits in conventional CMOS technology and then merge the two systems on a single chip. However, such an approach requires complicated interconnects and packaging techniques leading to significant power overhead. This issue can be addressed by integrating the analog interface with the SOT-based computing and storage systems. For doing this, design and implementation of SOT-based analog circuits is needed. In this section, the possibility of implementation of two critical analog/mixed-signal blocks including sensor for sensing e.g. biomedical signals and analog-to-digital converter (ADC) to digitize the sensed signals, will be introduced.

### A. Analog-to-Digital Conversion

ADC is an essential block in all systems which communicate with analog world. Such block translates an analog signal to digital information that can be processed by digital system. In some applications, e.g. image sensors, ADC needs to be very compact mainly due to the fact that the pixel size restricts the available area for ADC. More restrictive area is attributed to this fact that CMOS scaling enables smaller footprint digital circuits that allows significant downscaling of sensor pitches. Therefore, in such sensors, compactness is added to traditional concerns of ADC design which are resolution, conversion speed and power consumption. However, in CMOS analog circuits, mainly due to large process variation and low supply voltage, pace of scaling has become significantly slower. All these facts make ADC design in CMOS technology a challenging task for compact systems. To deal with this issue, new technologies such as spintronic devices may be considered as a solution [128]. Because of advantages of spintronic devices, those are prospective candidates to be used in ADCs and can enable benefit of ultra-compact ADC (1000X smaller) compared with conventional CMOS flash ADCs [128]. In [129], a 3-bit flash ADC based on SOT is presented in which SOT-based MTJs are employed as current mode comparators. In addition, critical currents of MTJs are used as the reference currents of the ADC. To implement such configuration, a parallel MTJ multi-level cell (P-MLC) introduced in the section II-C of the paper can be employed. In this structure, dimensions of the HM underneath each MTJ is appropriately engineered to make reference current of each comparator as shown in Fig. 12(a). The SOT-MTJ with bigger HM has higher $J_c$, hence the corresponding comparator has higher reference current. In addition, such ADC is able to store the digital output inside MTJs, hence no additional memory array is needed for storing the ADC output.

In the ADC shown in Fig. 12(a), HM engineering is used to scale the current reference of each comparator. Another approach, voltage controlled magnetic anisotropy (VCMA), is used to scale the references shown in Fig. 12(b) [130]. In this structure, MTJs have contact in parallel with a uniform HM underneath all MTJs. In addition, a transistor is connected to the third terminal of each MTJ while the source contact of each transistor is differently biased. To generate different bias voltages, a reference voltage is connected to a resistive ladder. By using this configuration, the transistors generate different assist STT currents passing through each MTJ. It means the critical current that SOT must overcome to enable switching is different in each MTJ. In other words, the critical current of MTJ is reduced proportional to the value of the assist STT current. In Fig. 12(b), the MTJ on the left side has the lowest $J_c$ because the source terminal of its transistor is biased by higher voltage. On the other hand, the MTJ on the right side has higher $J_c$ mainly due to lower bias voltage and STT current. This ADC has all the advantages of the structure in Fig. 12(a), while it may have easier fabrication process at the cost of more area overhead mainly due to using resistive ladder and transistors.

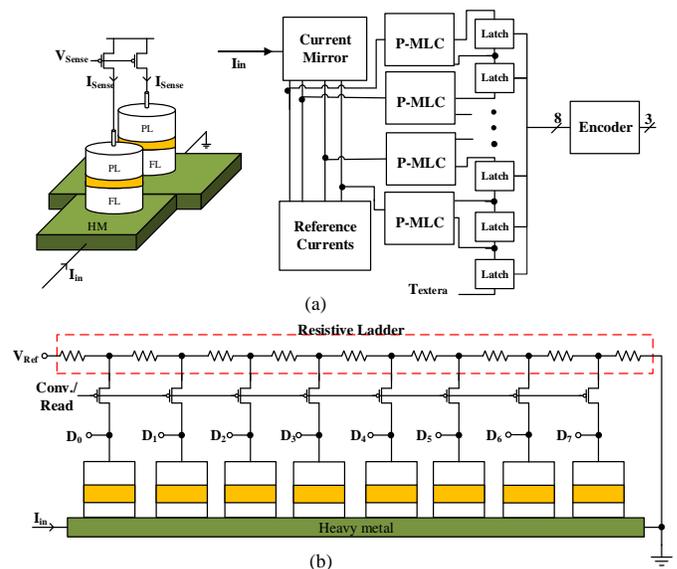

Fig. 12. (a) P-MLC and the realized ADC; P-MLC is used as a current comparator; reference currents are distinguished by engineering of HM (b) SHE-based 3-bit ADC; reference currents are distinguished by injecting different STT currents.

### B. Sensors

Detection and sensing of weak signals with high resolution through a contactless methodology is a crucial task in different applications such as biomedical applications, industrial electronics and non-destructive tests [131]. Spintronic giant magnetoresistive (GMR) and TMR sensors have been widely used in various applications [98], [132, 133]. The spin-based oscillators are able to sense weak signals by converting them into microwave signals. Generated microwave signals can be wirelessly transmitted to a coupled receiving micro-coil. Fig. 13 shows the schematic of the nano-oscillator based devices for different sensing applications. STNO has been reported as a wireless sensor for detecting of local currents, and its current sensitivity can be improved by the use of devices with lower threshold current [134].

In magnetic field sensing applications, comparing to the conventional magneto-resistive sensing devices, which are sensitive to signal amplitude, the spin oscillator-based magnetic sensors' performance relies on measuring the frequency of the oscillator. Authors in [135] have designed a STNO for maximizing performance carrying out macrospin simulations. Transition in magnetic media causes the changes in the direction of the applied magnetic field to the STNO, which can be sensed by the FL and this can lead to variation of the oscillator frequency. In SHNO device, the drive current is not coupled to the sensing bias across the MTJ. Thus, in these devices drive and sense path optimization can be carried out independently. Furthermore, the possibility to control the signal's amplitude and frequency independently in 3-terminal SHNOs make them promising devices for wireless sensing [68].

## VII. OTHER EMERGING APPLICATIONS

### A. Wireless Communication Systems

One of the main issues in wireless communication is to design a very compact, low phase-noise and wideband VCO. As discussed previously, SHNOs can generate signals in the range of GHz and above through different mechanisms such as precession of spins in the free magnetic e.g. microwave and RF-communication. These oscillators offer a high tunability and extremely smaller footprints that makes them superior compared to the LC tanks that are used in RF communication.

- ### RF and Microwave communication

Due to the possibility of generating high frequency signals from spin-based nano-oscillators, they have found some interests within the communication society. One of such applications, is to implement oscillators. Although this offers many advantages, it is still not possible to implement proper modulation due to its still low output power, although it can be improved by some techniques e.g. synchronization, as well as poor signal purity, nonlinearity in frequency-amplitude and mod hoping. In a work done by Choi et. al [136], a binary amplitude shift keying (ASK) was proposed and successfully tested for 1 meter between the transmitter and receiver with a data-rate of 5Mb/s for STNOs. Another work done by Oh *et. al.* proposes the use of On-Off-Shift-Keying (OOK) with a data rate of 400 kbps at the distance of 10mm and the power consumption of 3 mw per STNO including logic control circuitry [137]. Although not many works have been done using SOT devices, the same approach can be applied for them as they offer better scalability compared to the STNOs.

- ### THz communication

Most of the work done so far are focusing on the development of oscillators using SHNOs at GHz range. Recently, some works propose the replacement of the ferromagnetic layer with an antiferromagnetic layer enabling the possibility of emitting terahertz signal [138, 139], i.e. typically frequency range of $100\,\text{GHz-30 THz}$. Given such possibility, many applications can be emerged such as THz communication through. This research field is very new and many aspects are even untapped. To enable such possibility, we believe that much higher output power either through synchronization of many oscillators, novel devices or other approaches are required to be sensed from the receiving side without loss of the communicated data.

### B. Brain Interfacing Systems

With the great importance of development of novel interfacing circuits with smaller area, lower power and higher performance, especially for wearable or implantable applications, emerging technologies with such potential such as spintronics, have received a huge attention to develop novel devices e.g. low power sensors for Magnetoencephalography (MEG) [140], Electromyography (EMG) [141] and many other applications. Development of implantable NCS (e.g. based on spiking neural networks) in the brain for closed-loop systems can be the next wave of the implantable devices. Within this, in HERMES [142] project funded by H2020, we are exploring the possibility of implanting a memristor-CMOS NCS to heal the epileptic brain, and this can be extended to spintronic-CMOS-Memristor if the spintronic technology becomes more mature.

## VIII. FUTURE VISION AND EXISTING CHALLENGES

Although the field of spintronics has been researched thoroughly, only memories have been commercialized. As an example, emerging fields such as brain-inspired computing and implementation of neural networks using spin-based devices is very immature and requires years of research on how to fully implement such complicated architectures. Here, we discuss a few challenges that are facing the further development of in general, spin-based technologies.

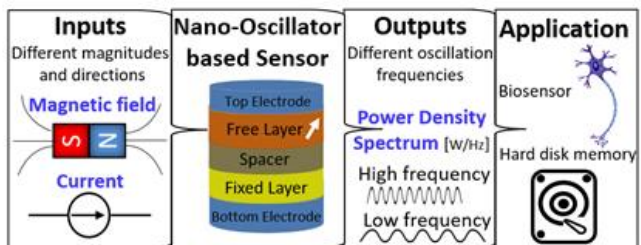

Figure 13. Spin Nano-Oscillator based sensing devices

- *Integration*

One main challenge is the integration of STNOs/SHNOs with CMOS due to their higher sensitivity for variation of the device properties. For oscillator application the viable resistance range is very tight and the stress on the tunnel barrier is high compared to memory applications. The main challenge is the additional roughness introduced by the lower CMOS layers. This roughness can be reduced by sacrificing device density and laterally separating spintronic devices from the underlying CMOS, but it has to be shown that to what degree oscillators can be realized on CMOS substrate. A second challenge is to adjust the deposition processes of the oscillators to reach a high density that is typical for CMOS devices.

- *CMOS-interfacing circuit*

Although for memory applications, different circuit techniques have been developed, other interfacing circuits for sensing, oscillation and computing are at their initial stages and almost no integrated solutions have been demonstrated. In this part, we consider different interfacing circuits proposed in literature for different applications while providing the current challenges within this area.

### A. Sensing Circuit for SOT-MRAMs

In the literature, the focus is on improving the switching of the MTJ in MRAMs. By scaling the technology, the switching improves inherently. On the other hand, the challenge moves to sensing operation mainly due to degradation of read margin and thermal stability. Due to these issues, conventional sensing circuits are not useful in scaled technology nodes. To deal with the mentioned obstacles, advanced sensing circuits and new sensing schemes have been developed.

In one of the sensing approaches so-called data-aware dynamic reference voltage, Reference voltage (VREF) is variable and changes based on the level of bitline Voltage (VBL). When data is "0", VREF increases while for "1" state, it decreases. This dynamic reference can significantly improve the read margin to overcome process variations, hence it improves the yield of sensing operation. Another approach to improve the yield is to cancel the offset voltage of the sense amplifier. Such approaches work based on latch-offset cancelation leading to less sensitivity to process variations and mismatches [143]. Another scheme, self-reference read, has also been developed, which compares the resistance of the MTJ with a reference resistance generated by the same MTJ [144]. Such techniques can be categorized mainly into destructive and non-destructive sensing schemes. In destructive schemes, original data is destroyed during the read operation, hence the data should be written back to the MTJ at the end of the read operation. On the other hand, non-destructive approaches preserve the original data during the read operation.

All the mentioned approaches have mainly been developed for STT-MRAMs but may also be used for conventional SOT-MRAMs. However, for other SOT-MRAMs e.g. multilevel cells, we may need to develop new sensing approaches.

### B. STNO/SHNO interfacing circuits (Read-out)

One of the main building blocks of any STNO/SHNO-based system is the sensing block which its duty is to read the current situation of STNOs/SHNOs. In case of switching approach, the sensing block can perform sensing either simultaneous [144, 145] or at a fixed time after applying the signal. The advantage of simultaneous sensing is higher speed and lower power consumption compared with using fixed times for sensing.

For oscillation's case, the power consumption and complexity of the sensing circuit increase as the output power of oscillation is very low (tens of nano-Watt). One approach to sense the oscillation is to amplify the signal and converting it to a DC signal (e.g. with a PD). Finally, the DC signal should be compared with a reference voltage by a comparator. Another approach to sense the oscillation of the signal is to use a reference STNO/SHNO and passing a reference current through it [146]. Then, through coupling the output signal of the reference and target STNO/SHNO, the oscillation of STNO/SHNO will be detected. The main drawback of this method is that the mismatch between reference STNO/SHNO and the target STNO/SHNO that can cause failure.

### IX. CONCLUSION:

In this paper, the spin-based devices and relevant research within this area with a focus on SOT devices was presented. Research within memory design assisted by different techniques including laser, strain and thermal, novel device approaches for building fast and dense memories, nano-oscillators and their challenges such as output power and read-out circuitries, and sensing applications were discussed. Furthermore, the possibility of using SOT devices for novel applications such as ADCs were discussed as well. In general, although SOT devices and in general spintronic devices have shown a high potential for non-memory applications, there are years of research required to prove the efficiency of using SOT for other applications such as neuromorphic computing or ADCs etc..

### X. ACKNOWLEDGMENT:

This work was supported by the European Union's Horizon 2020 Research and Innovation Programme under grant agreement No 824164 (HERMES) and 713481 (SPICE), and Marie Sklodowska-Curie Individual Fellowship 751089.